\documentclass[journal=jacsat,manuscript=article]{achemso}
\usepackage[utf8]{inputenc}
\usepackage[T1]{fontenc}
\usepackage[version=3]{mhchem}
\usepackage{soul}
\usepackage{bm}
\usepackage{threeparttable}
\usepackage{color}


\makeatletter
\let\@fnsymbol\@arabic
\makeatother

\title{Can We Learn the Energy of Sublimation of Ice from Water Clusters?}

\author{Joel M. Bowman}
\email{jmbowma@emory.edu}
\affiliation{Department of Chemistry and Cherry L. Emerson Center for Scientific Computation, Emory University, Atlanta, Georgia 30322, U.S.A.}

\author{Qi Yu}
\affiliation{Department of Chemistry, Fudan University, Shanghai, 200438, P.R. China}
\email{qi_yu@fudan.edu.cn}

\author{Chen Qu}
\author{Riccardo Conte}
\affiliation{Dipartimento di Chimica, Universit\`{a} degli Studi di Milano, via Golgi 19, 20133 Milano, Italy}

\author{Paul L. Houston}
\affiliation{Department of Chemistry and Chemical Biology, Cornell University, Ithaca, New York
14853, U.S.A. and Department of Chemistry and Biochemistry, Georgia Institute of
Technology, Atlanta, Georgia 30332, U.S.A}

\begin{document}

\clearpage
\begin{abstract}
This short paper reports a study of the electronic dissociation energies, $D_e$, of water clusters from direct \textit{ab initio} (mostly CCSD(T)) calculations and the q-AQUA and MB-pol potentials.  These clusters range in size from 6-25 monomers. These are all in very good agreement with each other, as shown in a recent Perspective by Herman and Xantheas. To the best of our knowledge, we present for the first time results for the D$_e$ per monomer.  To our surprise this quantity appears to be converging to a value close to 12 kcal/mol.  An estimate of 1.5 - 2 kcal/mol for the $\Delta$ZPE for these clusters puts the value of $D_0$ at 10 to 10.5 kcal/mol.  This value is remarkably (and probably fortuitously) close to the reported sublimation enthalpy of 10.2 kcal/mol at 10 K. However, given that these D$_e$ energies correspond to dissociation of the cluster to $N$ isolated monomers the interpretation of ``vaporization" of these ``solid" clusters is qualitatively reasonable.
\end{abstract}

%\flushbottom
%\thispagestyle{empty}
\clearpage
%\maketitle
%\newpage

Electronic dissociation energies, mostly at the CCSD(T) level of accuracy, for water clusters of $N$ = 2-11, 16, 17, 20 and 25 monomers were summarized in a recent Perspective by and Herman and Xantheas.\cite{Xantheas2023}  In addition the results obtained from ``explicit many-body potentials", were also reported and compared to the accurate \textit{ab initio} ones. These are the WHBB\cite{WHBB, WHBB1} MB-pol\cite{mbpol2b,mbpol3b}, q-AQUA\cite{q-AQUA} potential.  Here we consider just q-AQUA, MB-pol and the \textit{ab initio} results.  Fig. 1 shows a comparison of $D_e$ and as seen there is excellent agreement of the q-AQUA and MB-pol results with the \textit{ab initio} ones.  The lower panel shows the same comparison but for $D_e$ per monomer.  This quantity has not been considered before, at least to the best of our knowledge.  In any case, the $D_e$ per monomer appears to be converging, at the largest value of $N$ considered. To investigate this possible convergence with respect to $N$, we have plotted $D_e$ per monomer vs 1/$N$ in Fig. 2. As seen, the behavior is almost linear and so simple linear fits were done to each set of results.  The  extrapolated results for 1/$N$ equal to zero are shown in the figure. As seen, they are 11.5 (MB-pol), 11.9 (q-AQUA) and 12.0 (\textit{ab initio}) kcal/mol. When corrected with a reasonable estimate of zero-point energy differences (see below) these are remarkably (and probably fortuitously) close to the sublimation enthalpy of ice to vapor, which of course is a function of pressure and temperature.

\newpage
\begin{figure}[H]
\begin{center}
\includegraphics[width=0.8\textwidth]{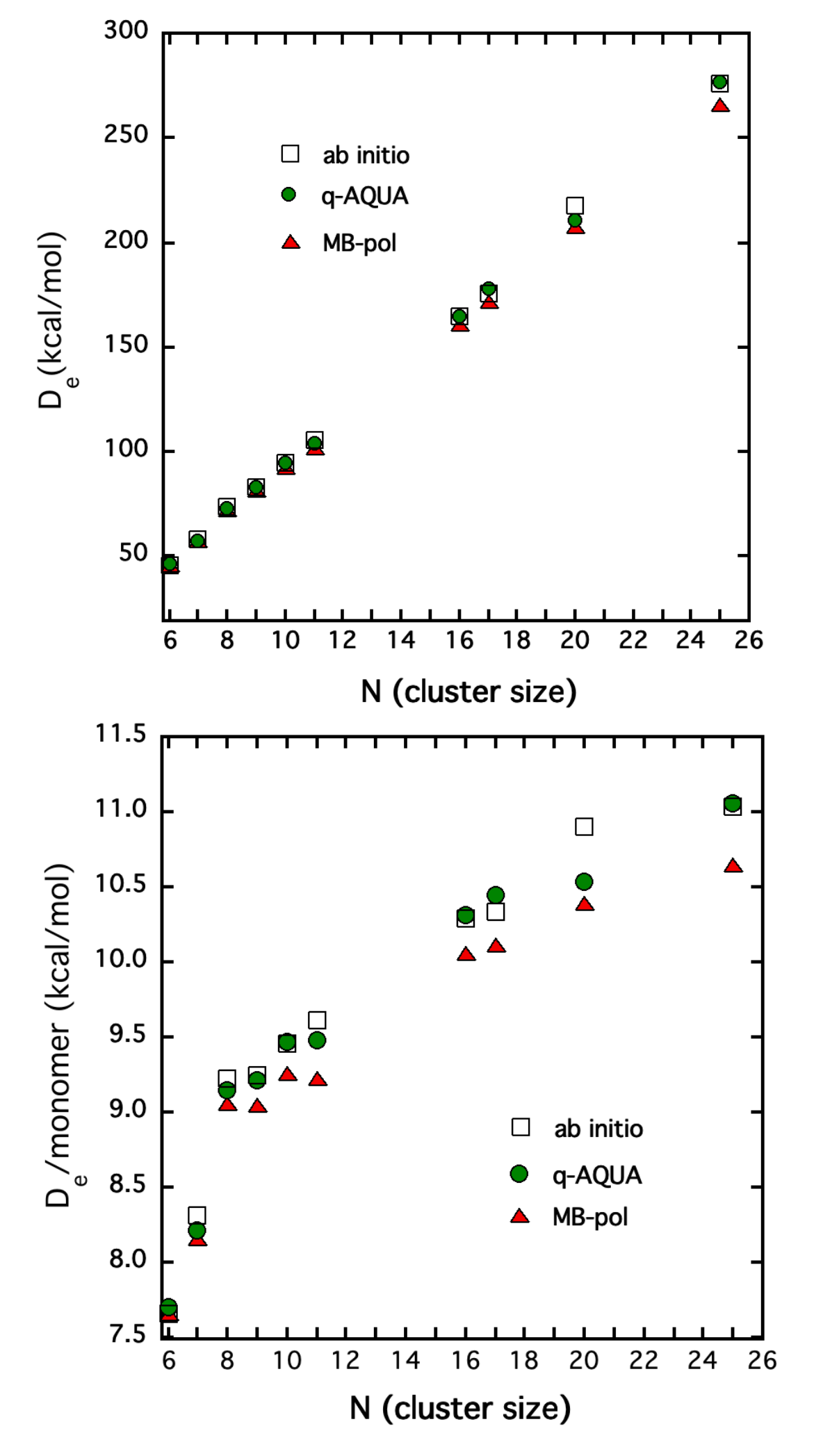}
\end{center}
\label{fig:diss energies}
\caption{Electronic dissociation energies, $D_e$, (upper panel) and  dissociation energies per monomer (lower panel) vs cluster size N.}
\end{figure}

\begin{figure}[H]
\begin{center}
\includegraphics[width=1.0\textwidth]{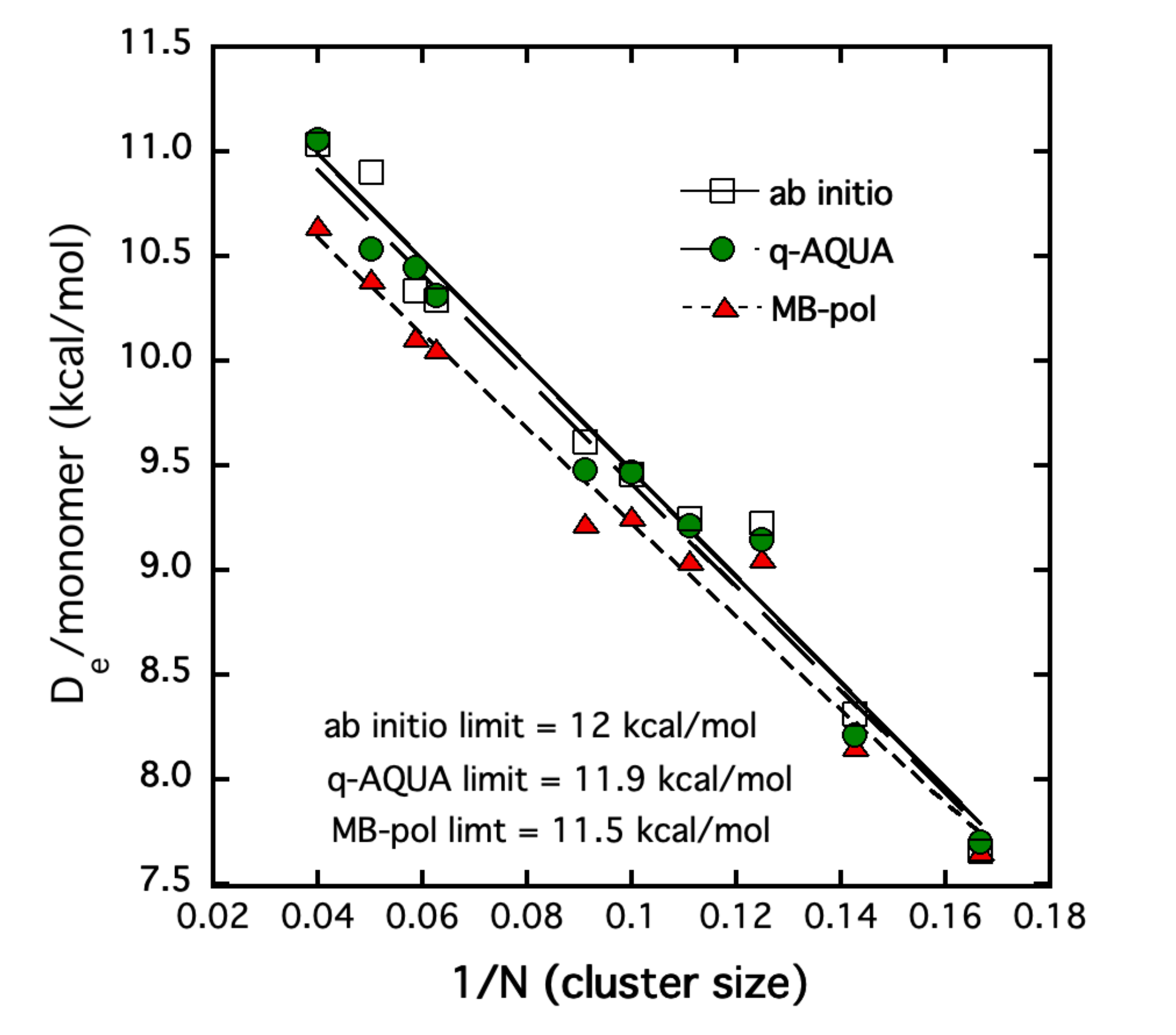}
\end{center}
\label{fig:diss 1/N}
\caption{Electronic dissociation energies, $D_e$, vs 1/N. Extrapolated limit values for 1/N equal zero are also given for indicated source. }
\end{figure}

Considering now zero-point energy and the difference between $D_e$ and $D_0$, which is just $D_e$ + $\Delta$ZPE.  There have not been systematic \textit{ab initio} studies of this for clusters as large as the ones considered here and in ref. \citenum{Xantheas2023}.  Some scattering of results in that reference, but not at the CCSD(T) level of theory, are given.   A systematic study of this can be made using the Supporting Information of the paper reporting the WHBB potential\cite{WHBB1}.  There normal mode frequencies using that potential were given for clusters up to $N$ = 22. Based on those and the known harmonic frequencies of the Partidge-Schwenke potential\cite{PS} we arrive at an estimate of around 2 kcal/mol $\Delta$ZPE for the clusters.  However, the trend of  $\Delta$ZPE with N is not smooth and clearly more study of this is needed.  In any case, based on that analysis the estimate for $D_0$/monomer would be value close to 10-10.5 kcal/mol, which is remarkably (and probably fortuitously) close to the thermodynamic enthalpy of sublimation,which at 10 and 273 K is 10.2 and 10.8 kcal/mol, respectively.\cite{sub2007}

\bibliography{ref}

\section{Data availability}
The data generated and used in this study are available at upon request to the authors.

\section*{Acknowledgment}
Q.Y. thanks Fudan University for the start-up funding. J.M.B. acknowledges support from NASA grant (80NSSC22K1167). D.H.Z. acknowledges the support from the National Natural Science Foundation of China (grant no. 22288201). R.C. thanks Universit\`{a} degli Studi di Milano for financial support under grant PSR2022\_DIP\_005\_PI\_RCONT.

\end{document}

\flushbottom
\maketitle
\clearpage

\begin{figure}[H]
\begin{center}
\includegraphics[width=0.4\textwidth]{Figures/SI_learning_curve.pdf}
\end{center}
\caption*{Supplementary Figure 1: Example of training RMSEs (meV/atom), relative to the MB-pol values of the potential energy, as a function of the number of training steps.}
\end{figure}

\begin{figure}[H]
\begin{center}
\includegraphics[width=0.4\textwidth]{Figures/SI_RDF.pdf}
\end{center}
\caption*{Supplementary Figure 2: OO, OH, and HH radial distribution function for liquid water at 298 K from classical molecular dynamics simulations on the MerNN model. The experimental data are taken from Refs \citenum{Skinner2013,Skinner2014}.}
\end{figure}

\newpage

\begin{figure}[H]
\begin{center}
\includegraphics[width=0.7\textwidth]{Figures/SI_MSD.pdf}
\end{center}
\caption*{Supplementary Figure 3: Mean square displacements as a function of time for liquid water at various temperatures using the MerNN model.}
\end{figure}

\begin{figure}[H]
\begin{center}
\includegraphics[width=0.7\textwidth]{Figures/SI_long_range.pdf}
\end{center}
\caption*{Supplementary Figure 4: OO radial distribution function for liquid water at 298 K from classical MD simulations using different MonoNet models. The MerNN-short is the model trained from a short cut-off (R$_\text{c}$=9.0 \AA) and used in the main text. The MerNN-long is the long-range model trained using a long cut-off (R$_\text{c}$=15.0 \AA). The experimental data are taken from Ref. \citenum{Skinner2013,Skinner2014}.}
\end{figure}

\bibliography{ref}